\documentclass[twocolumn,showpacs,amsmath,amssymb]{revtex4}
\usepackage{graphicx}

\begin{document}
\title{Chiral molecules split light: Reflection and refraction in a chiral liquid}
\author{Ambarish Ghosh}
\affiliation{The Rowland Institute at Harvard, Harvard University,
Cambridge, Massachusetts 02142}
\author{Peer Fischer}
\affiliation{The Rowland Institute at Harvard, Harvard University,
Cambridge, Massachusetts 02142}

\begin{abstract}
A light beam changes direction as it enters a liquid at an angle
from another medium, such as air. Should the liquid contain
molecules that lack mirror symmetry, then it has been predicted by
Fresnel that the light beam will not only change direction, but
will actually split into two separate beams with a small
difference in the respective angles of refraction. Here we report
the observation of this phenomenon. We also demonstrate that the
angle of reflection does not equal the angle of incidence in a
chiral medium. Unlike conventional optical rotation, which depends
on the path-length through the sample, the reported reflection and
refraction phenomena arise within a few wavelengths at the
interface and thereby suggest a new approach to polarimetry that
can be used in microfluidic volumes.
\end{abstract}
\maketitle

Liquids composed of randomly oriented molecules are, in the
absence of an external influence, isotropic, and are generally
described by a single scalar refractive index. Optically active
liquids, that is, chiral liquids which have the ability to rotate
the polarization vector of light, are an important exception.
These are characterized by two refractive indices, one for left-
$(-)$ and one for right- $(+)$ circularly polarized radiation.
Linearly polarized light may be regarded as a coherent
superposition of left- and right-circularly polarized waves of
equal amplitude, and Fresnel's theory of optical rotation shows
that a difference in the respective indices of refraction causes
the waves to acquire different phases as they propagate through
the liquid and the polarization vector to rotate
\cite{Fresnel1825,Barron}. The significance of optical rotation
lies in the fact that it is a means to distinguish the two mirror
image forms (enantiomers) of a chiral molecule. Most biologically
important molecules are chiral and optical rotation is a well
established analytical technique used to determine their absolute
stereochemical configuration (handedness) in solution. Optical
activity in liquids should, however, not only manifest itself
through optical rotation in transmission, but should also be
observable in reflection and refraction -- as we demonstrate in
this Letter.

\begin{figure}[t]
\includegraphics[scale=0.41]{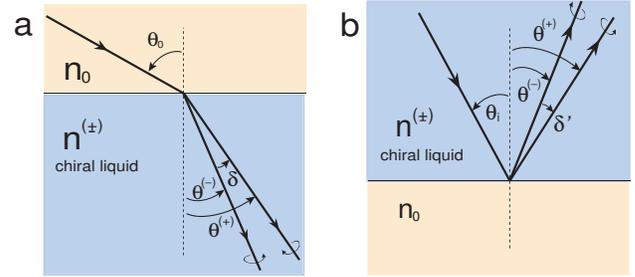}
\caption{\label{fig1} (color online) (a) Refraction geometry at an
achiral/chiral interface. A wave incident from an isotropic
achiral medium (angle of incidence $\theta_0$) is split into its
circularly polarized wave components in the chiral medium. In an
achiral medium the angle of reflection (not shown) does not depend
on the polarization of the light and is therefore not sensitive to
chirality. However, Silverman {\it et al.} showed that the
specular reflection off a chiral surface can exhibit differences
in intensity for the two circularly polarized components that
depend on the circular dichroism (differential absorption) in the
chiral medium \cite{Silverman1992}. Reflection geometry at a
chiral/achiral interface {\em in} a chiral medium. A wave incident
from a chiral medium is reflected at the interface and is split
into its circularly polarized wave components. The transmitted ray
(not shown) splits similar to the scheme shown in a).}
\end{figure}
If one considers the refraction of light at a boundary formed by a
chiral and an achiral isotropic medium, as shown in
Fig.\ref{fig1}a, then the circularly polarized components must
independently obey Snell's law:
\begin{equation}
n_0 \sin\theta_0 = n^{(\pm)} \sin\theta^{(\pm)} \; .
\end{equation}
Circularly polarized light, incident with angle of incidence
$\theta_0$ from the achiral medium characterized by the
polarization independent refractive index $n_0$, will thus refract
with angles of refraction $\theta^{(-)}$ and $\theta^{(+)}$,
depending on whether it is, respectively, left- or
right-circularly polarized
\cite{Fresnel,Fresnel1825,Lowry,Silverman1986}. Similarly, if
unpolarized or linearly polarized light is incident from the
achiral medium, then the light will split into two beams, one
left- and the other right-circularly polarized
\cite{Fresnel,Ditchburn1991}.

In order to verify that a linearly polarized light beam does split
into its left- and right-circularly polarized components as shown
in Fig.\ref{fig1}a, we imaged the separation between the two
refracted beams on a CCD camera. Given that the difference between
the refractive indices is small, the angular divergence $\delta=
\theta^{(+)}- \theta^{(-)}$ between the refracted beams in
Fig.\ref{fig1}a can be written as
\begin{equation}
\label{refraction} \delta \approx \frac{\tan\theta}{n} \left(
n^{(-)}-n^{(+)}\right) \; ,
\end{equation}
where $n=(n^{(-)}-n^{(+)})/2$ and where $\theta$ is the average of
the two angles of refraction. The splitting needs to be large
enough for it to be resolved on the detector. Fresnel showed that
multiple-refraction at non-parallel interfaces can be used to
increase the net divergence \cite{Fresnel,Fresnel1825}.
\begin{figure}
\includegraphics[scale=0.7]{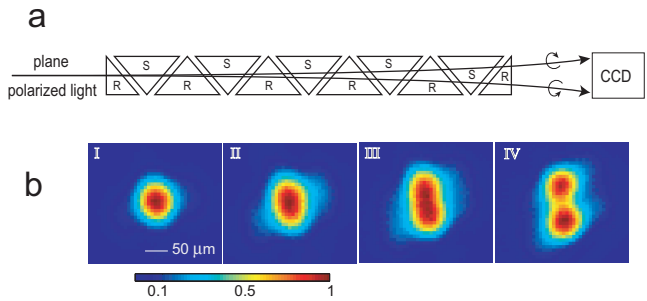}
\caption{\label{fig2} (color online) (a) Schematic of the
experimental geometry used to image double refraction. A plane
polarized light ray is incident onto a stack of prismatic
cuvettes, which are alternately filled with the R- and S-
enantiomer of a chiral liquid. The angular divergence increases by
$\sim2\delta$  at each R$\setminus$S or S/R interface, since both
the order of the enantiomers and the orientation of the interface
changes from one cuvette to the next. The beam would not split in
an achiral liquid. (b) Intensity plots of a laser with Gaussian
beam profile recorded on a CCD camera (4.6 micron pixel size). The
images correspond to a passage through 8, 12, 16, and 20
interfaces, (labeled I-IV respectively). Image IV clearly shows
two well separated Gaussian beam profiles that are right- (upper)
and left- (lower) circularly polarized.}
\end{figure}
Our experimental geometry, depicted in Fig.\ref{fig2}a, makes use
of this amplification scheme, and is similar to the one originally
devised by Fresnel to image double refraction in quartz crystals
\cite{Lowry,Fresnel,Fresnel1825}. In Figure \ref{fig2}b it is seen
that the splitting of a 405 nm diode laser beam, traversing
prismatic cuvettes filled with optically active solutions of
limonene, increases with the number of interfaces and becomes
clearly resolvable on a CCD camera. We verified that the two beams
in Fig.\ref{fig2}b IV are indeed circularly polarized and that
they are of opposite circularity. The refracted beams are of equal
intensity and are coherent as the incident light is plane
polarized. Should the incident light be unpolarized, then the two
circularly polarized refracted beams have no fixed phase-relation
\cite{Ditchburn1991}.

Multiple interfaces amplify the double-refraction effects, since
the angular deviation increases with each interface, but they are
not required in order to measure the chirality-induced splitting
of a light beam, which may also be observed in reflection. In the
case of reflection inside a liquid, as shown in Fig.\ref{fig1}b,
the angles can be obtained by considering the boundary conditions
of the electromagnetic waves at the interface. The latter requires
that the wavevectors of the incident $\vec{k}_i$ and the reflected
beam $\vec{k}_r$ have the same tangential components along the
interface, i.e. satisfy $k_i \sin\theta_i = k_r \sin\theta_r$,
where the angle of reflection (measured from the interface normal)
is $\theta_r$ , and where $k_r=2 \pi n_r/\lambda$ and $n_r$ is the
refractive index the reflected beam experiences. $k_i$ is
correspondingly defined for the incident beam. An ordinary achiral
liquid is described by a single refractive index and it follows
that the law of reflection ($\theta_r=\theta_i$) holds regardless
of the polarization state of the light. Because a circularly
polarized wave undergoes a (partial) polarization reversal upon
reflection, the wavevectors for the incident and the reflected
wave in a chiral liquid are associated with different refractive
indices. Hence, in an optically active medium
$\theta_r\neq\theta_i$, and there are two reflected waves with
angles of reflection $\theta^{(-)}$ and $\theta^{(+)}$
\cite{Lakhtakia1989,Silverman1994,Lalov1997}. Their angular
divergence after reflection is given by
\begin{equation}
\label{reflection} \delta' \approx \frac{\tan\theta_i}{n} \left(
n^{(-)}-n^{(+)}\right) \; .
\end{equation}
In general, the circular polarization components will not fully
reverse their circularity upon reflection, but become elliptically
polarized. The degree of circular polarization, and hence the
fraction of the beam that doubly reflects, depends on the Fresnel
reflection coefficients for a given interface, and needs to be
accounted for in practice.

The divergence between the two circular components of a reflected
or refracted beam can be directly compared with the optical
rotation in radians developed by light at the wavelength $\lambda$
traversing a distance $d$ in a chiral liquid
\cite{Barron,Condon1937}:
\begin{equation}
\label{ORD} \alpha=\frac{\pi \, d}{\lambda} \left( n^{(-)} -
n^{(+)} \right),
\end{equation}
which is also a function of the circular birefringence $n^{(-)} -
n^{(+)}$. It is thus expected that $\delta/\lambda$ and
$\delta'/\lambda$ will exhibit the same dependence on wavelength
as optical rotation does, provided the dispersion of the average
refractive index of the chiral medium in Eqs. (\ref{refraction})
and (\ref{reflection}) has been accounted for.
\begin{figure}[h]
\hspace*{-0.5cm}
\includegraphics[scale=0.7]{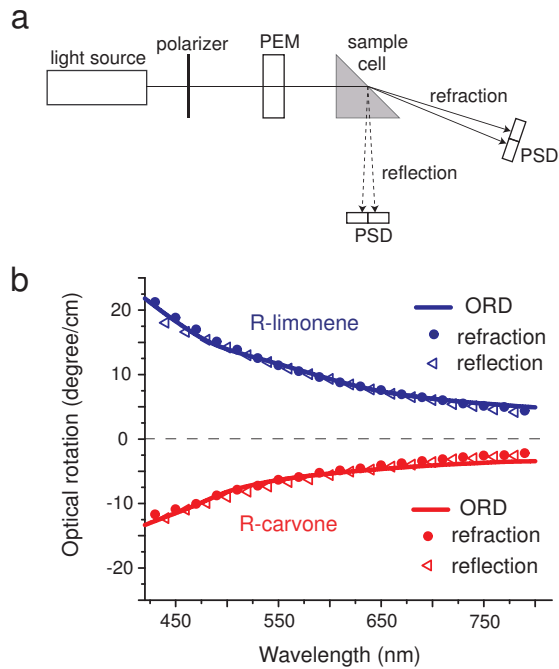}
\caption{\label{fig3} (color online) (a) Schematic of the
experimental geometry used for the reflection and refraction
measurements, where PSD denotes a position sensitive diode and PEM
a photoelastic modulator. In order to clearly separate reflection
and refraction phenomena, a $90^{\circ}$ prismatic cuvette holds
the chiral liquid, so that the doubly reflected beam exits the
cuvette without doubly refracting. (b) Optical rotation and
optical rotatory dispersion (ORD) data obtained from the
wavelength-dependence of the measured angular divergence in both
reflection (open triangles) and refraction (solid circles) at a
single achiral/chiral interface for two optically active liquids
R-($-$)-carvone (symbols in red) and R-($+$)-limonene (symbols in
blue). The data is seen to be in excellent agreement with
conventional optical rotation measurements obtained with a
spectropolarimeter (corresponding solid lines).}
\end{figure}
To test this, we passed light from a Xe arc lamp through a
monochromator and recorded $\delta$ and $\delta'$ as a function of
wavelength. The setup is schematically depicted in
Fig.\ref{fig3}a. A photoelastic modulator is used to modulate the
polarization of the light between left- and right-circularly
polarized at $\sim$50 kHz and the position of the light beam is
 synchronously recorded by a one-dimensional position sensitive
detector and a lockin amplifier. Figure \ref{fig3}b shows that the
angular divergence measurements obtained in reflection as well as
in refraction at a single interface between a chiral liquid and an
achiral medium, are in complete agreement with the conventionally
measured optical rotatory dispersion (ORD) spectrum. Our results
show that reflection and refraction can be used in place of
conventional optical rotation measurements, and their dispersion
correspondingly in place of ORD for structural and stereochemical
studies. The reflection data confirms that the `law of reflection'
does not necessarily hold in the presence of chiral molecules.
Similar results have been obtained with other optically active
liquids, such as pinene and aqueous glucose solutions.

\begin{figure}
\includegraphics[scale=0.7]{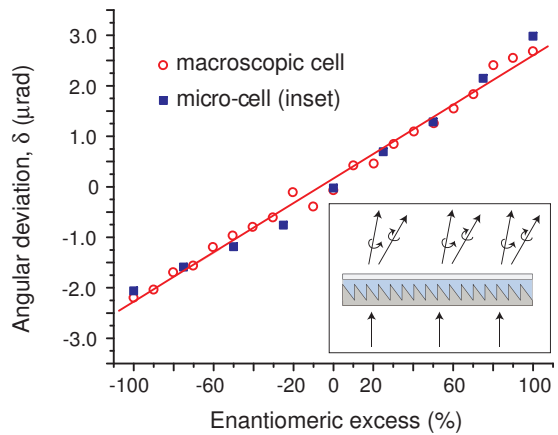}
\caption{\label{fig4} (color online) Angular divergence $\delta$,
measured in refraction, as a function of the concentration
difference between the two enantiomers of limonene, denoted by R
and S, respectively. The enantiomeric excess is defined as the
percent concentration difference between the enantiomers $\%[R]-
\%[S]$. The measurements are obtained at a single achiral/chiral
interface in two experimental geometries: a macroscopic cell (red
circles) and a micro-cell (blue squares), with average
path-lengths of $\sim$2 cm and $\sim$400 $\mu$m, respectively.
Each data point is measured with a standard deviation of $\sim$10
nanoradians and the straight line is a guide to the eye. The
experimental setup used in conjunction with the micro-cell is
shown in the inset: a thin film of liquid passes through part of a
commercial Fresnel lens (height of $45^{\circ}$ sawtooth-prisms
$\sim$700 $\mu$m), and a glass window separated by a 40 $\mu$m
spacer (not shown). The data sets have been obtained with a 488 nm
Ar+ and a 405 nm diode laser and have been normalized to 488 nm.}
\end{figure}
Away from the interface, the bulk of a liquid does not affect
reflection and refraction angles, and so it is interesting to ask
if it is indeed possible to miniaturize the sample cell and hence
the liquid volume without any loss of sensitivity. In doing so an
important aspect is the shape of the liquid cell. The two
circularly-polarized (cp) components will experience a small
parallel displacement upon traversing a thin slab geometry with
parallel front and exit windows, but will not experience any
change in their propagation directions. The parallel displacement
scales with the thickness of the fluid layer and this precludes
miniaturization. The reflection/refraction must therefore take
place in an asymmetric fashion, and in general, both phenomena
will contribute to the angular divergence between the cp beams. We
have constructed a micro-cell (inset Fig.\ref{fig4}) where the
liquid is confined asymmetrically by a part of a cylindrical
Fresnel lens. In Figure \ref{fig4} we compare the results from
double refraction measurements in the micro-cell where the average
path length through the chiral liquid is $\sim$400 $\mu$m with
results from a macroscopic cuvette where the light propagates
$\sim$2 cm through an optically active liquid. A polarimeter would
record an optical rotation in the micro-cell that is 50 times
smaller than that of the macroscopic cell, yet in Fig.\ref{fig4}
it is clearly seen that the measurements of the micro-cell and the
macroscopic cuvette give rise to the same angular divergences.
Since the circular birefringence of the chiral medium is
proportional to the concentration difference between the two
enantiomers, it follows that the angular divergence must be of
equal magnitude and of opposite sign as the solution changes its
handedness, and must be zero for the racemic (50:50) mixture, as
is observed for both cells.

An attractive feature of the interfacial refraction and reflection
phenomena is that the sample volume can be reduced further. If the
optical pathlength in the micro-cell is reduced to a few microns,
then already a modest \cite{resolution} angular resolution of 100
nanoradians (well within the current absolute accuracy of the
setup used in Fig.\ref{fig4}) will make it possible to observe
optical activities that are below the resolution limit of a
standard laboratory polarimeter, i.e. $<$ 0.001$^\circ$.
Miniaturization of optical components to micrometer dimensions is
expected to give rise to diffraction phenomena. This immediately
raises the interesting question whether the circular differential
reflection and refraction phenomena of this Letter can also be
observed in diffraction. Under normal incidence, the angular
position of the diffracted spot at order $m$ for a diffraction
grating with a groove spacing $D$ is given by $D \sin\beta=(m
\lambda)/n$, where $n$ is the refractive index of the light in the
medium where the diffraction grating is placed. Since the
refractive indices of right and left circularly polarized light
are different in an optically active medium ($n^{(\pm)}$), the
angular positions of the diffracted spots are also expected to
differ. We have made measurements with reflection gratings and
found that a chiral diffraction effect exists, such that left- and
right-circularly polarized components diffract with different
angles of diffraction \cite{tbp}:
\begin{equation}
D \sin\beta^{(\pm)}=(m \lambda)/n^{(\pm)} \; .
\end{equation}
We believe that these findings together with our observation of
double refraction and reflection in chiral liquids will open new
possibilities in the detection of optically active molecules in
small sample geometries such as in thin films or microfluidic
devices.

The unique optical properties of chiral media have also been
considered as components in electromagnetic coatings and microwave
devices \cite{Cory1995}, and more recently, in connection with
negative index media \cite{Pendry2004,Monzon2005}. Artificial
structures such as chiral helices have been used to make optically
active materials that give rise to relatively much larger optical
rotations at microwave frequencies
\cite{Lindmann1920,Tinoco1957,Guerin1994}. The double refraction
and reflection phenomena that we have observed in liquids are
therefore expected to give rise to correspondingly much larger
angular divergences and splittings in chiral metamaterials.

Finally, it is interesting to note that a magnetic field renders
any medium optically active, for a linearly polarized light wave
propagating along the direction of the field will experience
optical rotation. It thus follows, that the Faraday effect can
also be observed in reflection and refraction \cite{tbp} --
similar to the chiral phenomena reported in this Letter.

The authors thank Drs. M. Burns, Y.-F. Chen, and J.-M. Fournier
for helpful comments, J. Park for the loan of a diode laser, and
D. Rogers for the construction of a cuvette holder. Funding
through the Rowland Junior Fellows program is gratefully
acknowledged.


\begin{thebibliography}{19}
\expandafter\ifx\csname
natexlab\endcsname\relax\def\natexlab#1{#1}\fi
\expandafter\ifx\csname bibnamefont\endcsname\relax
  \def\bibnamefont#1{#1}\fi
\expandafter\ifx\csname bibfnamefont\endcsname\relax
  \def\bibfnamefont#1{#1}\fi
\expandafter\ifx\csname citenamefont\endcsname\relax
  \def\citenamefont#1{#1}\fi
\expandafter\ifx\csname url\endcsname\relax
  \def\url#1{\texttt{#1}}\fi
\expandafter\ifx\csname
urlprefix\endcsname\relax\def\urlprefix{URL }\fi
\providecommand{\bibinfo}[2]{#2}
\providecommand{\eprint}[2][]{\url{#2}}

\bibitem[{\citenamefont{Fresnel}(1825)}]{Fresnel1825}
\bibinfo{author}{\bibfnamefont{A.}~\bibnamefont{Fresnel}},
  \bibinfo{journal}{Ann. Chim. Phys.} \textbf{\bibinfo{volume}{28}},
  \bibinfo{pages}{147} (\bibinfo{year}{1825}).

\bibitem[{\citenamefont{Barron}(2004)}]{Barron}
\bibinfo{author}{\bibfnamefont{L.~D.} \bibnamefont{Barron}},
  \emph{\bibinfo{title}{Molecular light scattering and optical activity}}
  (\bibinfo{publisher}{Cambridge University Press}, \bibinfo{year}{2004}),
  \bibinfo{edition}{2nd} ed.

\bibitem[{\citenamefont{Silverman et~al.}(1992)\citenamefont{Silverman, Badoz,
  and Briat}}]{Silverman1992}
\bibinfo{author}{\bibfnamefont{M.~P.} \bibnamefont{Silverman}},
  \bibinfo{author}{\bibfnamefont{J.}~\bibnamefont{Badoz}}, \bibnamefont{and}
  \bibinfo{author}{\bibfnamefont{B.}~\bibnamefont{Briat}},
  \bibinfo{journal}{Opt. Lett.} \textbf{\bibinfo{volume}{17}},
  \bibinfo{pages}{886} (\bibinfo{year}{1992}).

\bibitem[{\citenamefont{Fresnel}(1822)}]{Fresnel}
\bibinfo{author}{\bibfnamefont{A.~J.} \bibnamefont{Fresnel}}, in
  \emph{\bibinfo{booktitle}{{\OE}vres compl\`{e}tes d'Augustin Fresnel}},
  edited by \bibinfo{editor}{\bibfnamefont{H.~d.} \bibnamefont{S\'{e}narmont}},
  \bibinfo{editor}{\bibfnamefont{E.}~\bibnamefont{Verdet}}, \bibnamefont{and}
  \bibinfo{editor}{\bibfnamefont{L.}~\bibnamefont{Fresnel}}
  (\bibinfo{publisher}{Paris}, \bibinfo{year}{1822}), vol.~\bibinfo{volume}{1}.

\bibitem[{\citenamefont{Lowry}(1964)}]{Lowry}
\bibinfo{author}{\bibfnamefont{T.~M.} \bibnamefont{Lowry}},
  \emph{\bibinfo{title}{Optical rotatory power}} (\bibinfo{publisher}{Dover
  Publications}, \bibinfo{address}{New York}, \bibinfo{year}{1964}).

\bibitem[{\citenamefont{Silverman}(1986)}]{Silverman1986}
\bibinfo{author}{\bibfnamefont{M.~P.} \bibnamefont{Silverman}},
  \bibinfo{journal}{J. Opt. Soc. Am. A} \textbf{\bibinfo{volume}{3}},
  \bibinfo{pages}{830} (\bibinfo{year}{1986}).

\bibitem[{\citenamefont{Ditchburn}(1991)}]{Ditchburn1991}
\bibinfo{author}{\bibfnamefont{R.~W.} \bibnamefont{Ditchburn}},
  \emph{\bibinfo{title}{Light}} (\bibinfo{publisher}{Dover Publications},
  \bibinfo{address}{New York}, \bibinfo{year}{1991}).

\bibitem[{\citenamefont{Lakhtakia et~al.}(1989)\citenamefont{Lakhtakia,
  Varadan, and Varadan}}]{Lakhtakia1989}
\bibinfo{author}{\bibfnamefont{A.}~\bibnamefont{Lakhtakia}},
  \bibinfo{author}{\bibfnamefont{V.~V.} \bibnamefont{Varadan}},
  \bibnamefont{and} \bibinfo{author}{\bibfnamefont{V.~K.}
  \bibnamefont{Varadan}}, \bibinfo{journal}{J. Opt. Soc. Am. A}
  \textbf{\bibinfo{volume}{6}}, \bibinfo{pages}{23} (\bibinfo{year}{1989}).

\bibitem[{\citenamefont{Silverman and Badoz}(1994)}]{Silverman1994}
\bibinfo{author}{\bibfnamefont{M.~P.} \bibnamefont{Silverman}}
  \bibnamefont{and} \bibinfo{author}{\bibfnamefont{J.}~\bibnamefont{Badoz}},
  \bibinfo{journal}{J. Opt. Soc. Am. A} \textbf{\bibinfo{volume}{11}},
  \bibinfo{pages}{1894} (\bibinfo{year}{1994}).

\bibitem[{\citenamefont{Lalov and Georgieva}(1997)}]{Lalov1997}
\bibinfo{author}{\bibfnamefont{I.~J.} \bibnamefont{Lalov}} \bibnamefont{and}
  \bibinfo{author}{\bibfnamefont{E.~M.} \bibnamefont{Georgieva}},
  \bibinfo{journal}{J. Mod. Opt.} \textbf{\bibinfo{volume}{44}},
  \bibinfo{pages}{265} (\bibinfo{year}{1997}).

\bibitem[{\citenamefont{Condon}(1937)}]{Condon1937}
\bibinfo{author}{\bibfnamefont{E.~U.} \bibnamefont{Condon}},
  \bibinfo{journal}{Rev. Mod. Phys.} \textbf{\bibinfo{volume}{9}},
  \bibinfo{pages}{432} (\bibinfo{year}{1937}).

\bibitem[{res()}]{resolution}
\bibinfo{note}{We note that different detection schemes permit the resolution
  of 0.1 nanoradians, see for instance R.V. Jones, J. of Sci. Inst. {\bf 38} 37
  (1961).}

\bibitem[{tbp()}]{tbp}
\bibinfo{note}{A. Ghosh and P. Fischer, manuscript in preparation (2006).}

\bibitem[{\citenamefont{Cory}(1995)}]{Cory1995}
\bibinfo{author}{\bibfnamefont{H.}~\bibnamefont{Cory}}, \bibinfo{journal}{J.
  Electrom. Waves Appl.} \textbf{\bibinfo{volume}{9}}, \bibinfo{pages}{805}
  (\bibinfo{year}{1995}).

\bibitem[{\citenamefont{Pendry}(2004)}]{Pendry2004}
\bibinfo{author}{\bibfnamefont{J.~B.} \bibnamefont{Pendry}},
  \bibinfo{journal}{Science} \textbf{\bibinfo{volume}{306}},
  \bibinfo{pages}{1352} (\bibinfo{year}{2004}).

\bibitem[{\citenamefont{Monzon and Forester}(2005)}]{Monzon2005}
\bibinfo{author}{\bibfnamefont{C.}~\bibnamefont{Monzon}} \bibnamefont{and}
  \bibinfo{author}{\bibfnamefont{D.~W.} \bibnamefont{Forester}},
  \bibinfo{journal}{Phys. Rev. Lett.} \textbf{\bibinfo{volume}{95}},
  \bibinfo{pages}{123904} (\bibinfo{year}{2005}).

\bibitem[{\citenamefont{Lindmann}(1920)}]{Lindmann1920}
\bibinfo{author}{\bibfnamefont{K.~L.} \bibnamefont{Lindmann}},
  \bibinfo{journal}{Ann. Phys.} \textbf{\bibinfo{volume}{63}},
  \bibinfo{pages}{621} (\bibinfo{year}{1920}).

\bibitem[{\citenamefont{Tinoco and Freeman}(1957)}]{Tinoco1957}
\bibinfo{author}{\bibfnamefont{I.}~\bibnamefont{Tinoco}} \bibnamefont{and}
  \bibinfo{author}{\bibfnamefont{M.~P.} \bibnamefont{Freeman}},
  \bibinfo{journal}{J. Phys. Chem.} \textbf{\bibinfo{volume}{61}},
  \bibinfo{pages}{1196} (\bibinfo{year}{1957}).

\bibitem[{\citenamefont{Gu\'{e}rin}(1994)}]{Guerin1994}
\bibinfo{author}{\bibfnamefont{F.}~\bibnamefont{Gu\'{e}rin}},
  \bibinfo{journal}{PIER} \textbf{\bibinfo{volume}{9}}, \bibinfo{pages}{219}
  (\bibinfo{year}{1994}).

\end{thebibliography}

\end{document}